\documentclass[12pt]{aastex}                % Courbin
\usepackage{emulateapj5}                    % Courbin

\shorttitle{
Cosmic alignment towards the radio Einstein ring PKS~1830--211 ?}
\shortauthors{Courbin et~al.}

\begin{document}


\title{
Cosmic alignment towards the radio Einstein ring PKS~1830--211\thanks{} ?
}


\author{ F. Courbin\altaffilmark{2}} 
\affil{Institut  d'Astrophysique  et  de G\'eophysique,  
Universit\'e  de  Li\`ege, All\'ee du 6 Ao\^ut 17, \\ Sart Tilman 
(Bat. B5C), Li\`ege 1, Belgium,~Frederic.Courbin@ulg.ac.be}

\author{ G. Meylan\altaffilmark{3}}
\affil{  Space Telescope Science Institute,
         3700 San Martin Drive, 
         Baltimore, MD 21218, U.S.A.,~
         gmeylan@stsci.edu }

\author{ J.-P. Kneib\altaffilmark{}}
\affil{  Observatoire Midi-Pyr\'en\'ees, 
         Laboratoire d'Astrophysique, UMR5572, 
         14 Avenue Edouard Belin, \\
         31000 Toulouse, France,~
         kneib@ast.obs-mip.fr }

\author{ C. Lidman\altaffilmark{}}
\affil{  European  Southern Observatory,   
         Casilla 19001, Santiago 19, Chile,~ 
         clidman@eso.org}
 

\altaffiltext{1}{Based on  observations made with  the NASA/ESA Hubble
        Space Telescope,  obtained from the Data Archive  at the Space
        Telescope  Science   Institute,  which  is   operated  by  the
        Association of  Universities for Research  in Astronomy, Inc.,
        under NASA contract NAS~5-26555.   The HST data used here come
        from the  archives related to  the programs \# 7495,  \# 8804,
        and \#  9133 (CASTLES). 
        This  study is also  based on  observations made
        with the NOAO Gemini-North  Telescope; NOAO is operated by the
        Association of Universities  for Research in Astronomy (AURA),
        Inc.  under cooperative  agreement with  the  National Science
        Foundation.}

\altaffiltext{2}{Pontificia Universidad
Cat\'olica de Chile, Departamento de Astronomia y Astrofisica, Casilla
306, Santiago 22, Chile.}

%\altaffiltext{3}{GEPI, Observatoire de Paris, Place
%Jules Janssen 5, F-92195, Meudon C\'edex, France.}

\altaffiltext{3}{Affiliated  with  the  Astrophysics Division  of  the
                 European   Space   Agency,   ESTEC,  Noordwijk,   The
                 Netherlands. }

%\altaffiltext{3}{present address:  Center for Astrophysics,  60 Garden
%                 Street, Cambridge, MA 02138}

%%%%%%%%%%%%%%%%%%%%%%%%%%%%%%%%%%%%%%%%%%%%%%%%%%%%%%%%%%%%%%%%%%%%%%


\begin{abstract}
Optical and  near-IR Hubble Space Telescope  and Gemini-North adaptive
optics  images, further  improved through  deconvolution, are  used to
explore  the gravitationally lensed  radio source  PKS~1830--211.  The
line of sight to the quasar at $z=2.507$ appears to be very busy, with
the presence, within  0.5\arcsec\, from the source of:  (i) a possible
galactic main-sequence  star, (ii) a faint red  lensing galaxy visible
only in  $H$-band and (iii) a  new object whose  colors and morphology
match those  of an almost face-on  spiral.  The $V-I$  color and faint
$I$ magnitude  of the  latter suggest that  it is associated  with the
molecular absorber seen towards PKS~1830--211, at $z=0.89$ rather than
with the $z=0.19$  HI absorber previously reported in  the spectrum of
PKS~1830--211.  While this discovery  might ease the interpretation of
the observed absorption lines, it also complicates the modeling of the
lensing potential  well, hence decreasing  the interest in  using this
system as a mean to measure  $H_0$ through the time delay between the
lensed images. This is the first  case of a quasar lensed by an almost
face-on spiral galaxy.
\end{abstract}



%%%%%%%%%%%%%%%%%%%%%%%%%%%%%%%%%%%%%%%%%%%%%%%%%%%%%%%%%%%%%%%%%%%%%%


\section{Historical context}

This paper presents a new step in the long series of studies trying to
unveil   the   properties  of   the   gravitationally  lensed   quasar
PKS~1830--211.   The object  has a  heavily populated  line  of sight,
crowded  by  galactic  and   extragalactic  objects.   We  start  this
introduction with  a few  highlights of what  has been  a particularly
slow discovery process.

The radio source  PKS~1830--211 was first detected as  a single source
in the Parkes survey (Shimmins et al. 1969) and soon recognized as one
of  the brightest  sources  at centimeter  wavelengths.  About  twenty
years later, high  spatial resolution VLA observations at  1.5, 5, and
15~GHz unveiled  the double structure of  this otherwise flat-spectrum
source  (Rao et al.  1988).  These  authors immediately  mentioned 
gravitational  lensing   as  the   best  qualitative
explanation for the presence  of the two components, their separation,
their  flux ratios, and  their almost  identical substructures  with a
point  inversion  symmetry  with   respect  to  each  other.   Further
high-resolution  VLA  observations at  5,  15  and 22.5~GHz  provided
enough  information  to  allow  theoretical modeling  of  the  lensing
effect, in spite  of the unknown redshifts of both  the source and the
lens (Subrahmanyan et al. 1990).

Deeper VLBI, Merlin  and VLA observations, at 2.3,  1.7 and 8.4 GHz,
respectively,  showed   an  unusual  elliptical   ring-like  structure
connecting  the two  brighter  components (Jauncey et al. 1991).  
This  further favored  the
gravitational lens explanation,  specifically, an Einstein ring formed
by  the  gravitational imaging  of  a  background  radio source  by  a
foreground mass concentration.  Extensive theoretical 
modeling studies which reproduced successfully
the ring and the two bright components were soon published (Kochanek
\& Narayan 1992, Nair et~al.  1993).

Since the line of sight to PKS~1830--211 passes the galactic bulge
(galactic longitude $l = 12.2\deg$ and galactic latitude $b =
-5.7\deg$), earlier attempts to identify the source and the lens
failed because of source confusion, faintness, and poor spatial
resolution. No visible counterpart was found down to $B$
$\simeq$ 23~mag.   A relatively bright object,  nearly coincident with
the  North-East   component  of  the  radio   source, was identified 
spectroscopically with the Palomar 200-inch telescope
as a foreground M star (Djorgovski  et al.  1992). Extensive  theoretical
modeling studies  which reproduced successfully  the ring and  the two
bright components were soon  published (Kochanek \& Narayan 1992, Nair
et~al.  1993).

Wiklind \& Combes (1996) identified 12 molecular absorption lines in a
millimeter spectrum of PKS~1830--211 taken with the 15-m SEST/ESO
telescope.  They inferred that the lines originate in the lensing
galaxy at $z$=0.89 and that the lens is a spiral, since the molecular
abundances are compatible with Milky Way values. 
Lovell et~al.   (1996) observed HI
absorption at  $z$ = 0.19 in  Parkes data, indicating  the presence of
another  galaxy  at this  redshift  and  pointing  towards a  possible
compound  lens (see also  Frye et~al.   1997).  The  $z$=0.89 redshift
value  was confirmed  by Mathur  \&  Nair (1997)  who observed  strong
absorption  features  in  their   X-ray  spectrum  acquired  with  the
Rosat/PSPC.  HI absorption is also  seen at $z$=0.89 (Chengalur et al.
1999).  Further observations  made at the 30-m IRAM  and 15-m SEST/ESO
telescopes confirmed  the presence  of two absorption  lines at  $z$ =
0.89,  one  corresponding  to  the  South-West  lensed  image  of  the
background   source,   and  the   other,   shifted   in  velocity   by
{--147~km\thinspace s$^{-1}$}, corresponding  to the North-East image.
This  implied  that the  background  radio  source  is situated  at  a
redshift of  about 3 and consistent  with the lensing  galaxy being an
early-type  spiral seen almost  face-on (Wiklind  \& Combes  1998; see
also Swift et~al.  2001).

At that stage,  models predicted time delays of the order  of one day to
several tens of days for reasonable range of source and lens redshifts
(Nair  et  al. 1992).   This expectation and the  fact  that
PKS~1830--211  varies regularly  on  time-scales of  months made  this
object an  ideal target, at that time,  for determining a time delay 
between the lensed components.
VLA observations at 8.4 and 15~GHz over a period of 13 months provided
a first estimate $\Delta t = 44 \pm 9$ days (Ommen et al. 1995).  More
recently,  ATCA observations  at 8.6~GHz  over a  period of  18 months
provided a second --- inconsistent ---  estimate $\Delta t = 26 \pm 5$
days  (Lovell et  al. 1998).   Since the  redshift of  the  source was
unknown and  the lensing galaxy  poorly parameterized, no  estimate of
$H_0$ was  obtained. Microlensing, as deduced from ASCA X-ray observations 
(Oshima et al. 2001), may complicate the computation of the time delay.

It was only with the advent of a powerful deconvolution method (Magain
et al.  1998) that  significant progress was  made in  identifying the
optical-infrared counterparts of  PKS~1830--211. Near-infrared $J$ and
$K$ images  obtained at  the 2.2-m MPI/ESO  telescope and $I$  and $K$
images with the Keck~I  telescope, all with sub-arcsecond seeing, were
deconvolved  by  Courbin et~al.   (1998).   Both  counterparts of  the
flat-spectrum core of  the radio source were searched  for. The M star
identified by Djorgovski et~al.  (1992) was clearly separated from the
bright  North-East component of  the quasar,  whose radio  and optical
positions were shown to be identical.  The South-West radio component,
with  similar radio flux,  was not  unambiguously identified  with the
optical  object, because of  mismatches between  the positions  of the
source in  the $I,J$,  and $K$  bands, and with  its radio  position as
well. This  object, much fainter in  $I,J$ and $K$ than  in the radio,
was possibly  identified with the lensing  galaxy alone or  a blend of
various objects. The source counterparts are very red with $I-K \sim 7$, 
which suggests strong
absorption from the Galaxy, the lensing galaxy or both.

The brightness of the source at near-infrared wavelengths enabled the
redshift of the source ($z$=2.507 $\pm$ 0.002) to be determined with
infra-red spectroscopy  (Lidman et~al. 1999).

Using   intrinsically  high  spatial   resolution  images
obtained with  the Hubble Space  Telescope in the near  infra-red ($I$
with  WFPC2/HST  and  $H,K$  with NIC2/HST),  Leh\'ar  et~al.   (2000)
confirmed and improved the  earlier quasar  component identifications proposed
by Courbin et~al.  (1998) with their ground-based deconvolved data. In
addition,  Leh\'ar  et~al.   (2000)  identified an  object  that  they
associated with the lensing galaxy  (lens G in Leh\'ar et~al. 2000 and
in the following).

In this paper  we present new results obtained  from the deconvolution
of deep imaging archive data  obtained with the Hubble Space Telescope
at  visible  and near-IR  wavelengths,  namely, WFPC2/F555W/F814W  and
NICMOS2/F160W/F205W.   In  addition,  we  use  some  new  ground-based
$K$-band  adaptive   optics  images  from   the  Gemini-North/Hokupa'a
Adaptive  Optics  (AO) system,  further  improved with  deconvolution.
These observations  in four broad-band  filters, viz., $V, I,  H$, and
$K$, allow us to detect  very clearly the  arms of a  spiral galaxy
between  the two quasar  images. In  addition to  this new  object, two
other objects are seen along the line of sight: (i) a red point source
that might be  either the bulge of the spiral or  a galactic star, and
(ii) a very faint and red object  seen only in $H$, at the position of
the object G previously reported by Leh\'ar et~al. (2000).

%%%%%%%%%%%%%%%%%%%%%%%%%%%%%%%%%%%%%%%%%%%%%%%%%%%%%%%%%%%%%%%%%%%%%%

\section{Observations}

\subsection{Hubble Space Telescope Imaging}

The Hubble  Space Telescope observations  presented in this  paper are
public and available  in the archives.  They were  obtained as part of
GO programs (PI: E. Falco with ID \# 7495, \# 8804, and \# 9133) known
as the CfA-Arizona Space Telescope LEns Survey (CASTLES), and aimed at
imaging  with the HST  all known  gravitationally lensed  quasars. The
near-IR data,  viz., WFPC2/F555W/F814W and  NIC2/F160W/205W, collected
by CASTLES for PKS~1830--211 were  used in their study of 10 two-image
gravitational lenses  (Leh\'ar et~al.  2000), where the observing 
details can be found.

We  re-analyse here  these  data, using  deconvolution techniques  for
further improvement.  We  take also advantage of the  fact that, since
the  publication of  Leh\'ar et~al.   (2000), much  deeper WFPC2/F814W
images were obtained (program ID \# 8804), corresponding to four times
1,200~sec,  in   addition  to  the  four   400-sec  exposures  already
published.   These additional  data  allow us  to  discover an  almost
face-on  spiral  galaxy  acting  as  a   gravitational  lens  on
PKS~1830--211  (see  below).   We  also  detect  this  galaxy  in  new
WFPC2/F555W data  (3 images, for a  total of 2,000 sec;  program ID \#
9133), however very close to the detection limit.

\subsection{Gemini/Hokupa'a Adaptive Optics Imaging}

The above  HST data are  supplemented by ground-based  $K$-band  
adaptive optics (AO) images of PKS~1830--211,  obtained during the  night of
2000  August 8$^{th}$,  with the  Hokupa'a instrument  mounted  on the
8.2-m  Gemini-North telescope at  Mauna Kea,  Hawaii. 
The raw data have excellent sampling (0.020\arcsec\, per pixel) and
fairly good image quality (FWHM = 0.15\arcsec). The 
 reduced data before deconvolution 
is presented in Figure \ref{fig1}, where
the  two  quasar images  are  seen, well  separated,  as  well as  the
$V~\sim~13.5$~mag  star  used  for  the  wavefront  correction,  about
7\arcsec\,  to  the North-East  of  PKS~1830--211.   Due  to the  high
airmass  of the object  (between 1.5  and 2.2)  and to  the relatively
faint guide star, the wavefront correction was not fully satisfactory,
with a final  seeing about twice as large as  the diffraction limit of
an 8.2-m telescope in the  $K$-band.  These images are rather shallow,
with a total exposure time  of 30 minutes, but they nevertheless allow
the measurement of the positions of the quasar images with an accuracy
better than could  be done with the HST data  alone.  No standard star
was observed, as the observations were performed through thin cirrus.

These data are the first images  of a gravitational lens taken with AO
on a 10-m class telescope.  The presence of a suitable wavefront guide
star and many PSF stars should allow one to improve the present result
if the observations can be repeated at a smaller airmass and with much
longer exposure time.  Given  the low declination of PKS~1830-211, the
Paranal Observatory is ideally located for such observations.

%%%%%%%%%%%%%%%%%%%%%%%%%%%%%%%%%%%%%%%%%%%%%%%%%%%%%%%%%%%%%%%%%%%%%%

\section{Astrometry --- Photometry}

All  the data  are deconvolved  with the  MCS  deconvolution algorithm
which  results in images  with improved  resolution and  sampling (see
Courbin et~al.   1998 for more  details about the application  of this
method). The spatial resolution  of the final images is 0.046\arcsec\,
for   the   WFPC2/F555W/F814W  data,   0.075\arcsec\,   for  the   HST
NIC2/F160W/F205W  near-IR  data,  and  0.020\arcsec\, for  the  Gemini
$K$-band data.

Figures  \ref{fig2},  \ref{fig3}, and  \ref{fig4}  show the  resulting
deconvolved images. In all cases  but for the F205W filter, the result
of  the  deconvolution process  is  good,  as  several PSF  stars  are
available in the immediate vicinity of the deconvolved field.

Since the objects in the  vicinity of PKS~1830-211 have very different
colours, astrometry is  performed in all available bands,  taking as a
reference coordinate system the best and deepest data available, i.e.,
the HST F814W image.  The transformation between the astrometry of the
different bands into the coordinate  system of the F814W image is done
by using  stars seen  simultaneously in all  filters.  Tables 1  and 2
gives the photometric and astrometric results for all selected objects
(see Section~4 for the meaning of  the object labeling that we keep as
in Leh\'ar et~al.  2000).

The photometry and astrometry  are always performed on the deconvolved
images  in order  to avoid  mutual contamination  of the  extended and
unresolved objects. We used the  same zero points as in Leh\'ar et~al.
(2000).  The  magnitude of extended objects are  measured in apertures
of  0.5\arcsec\, in  diameter.   As PKS~1830-211  is  situated at  low
galactic  latitude,  we have  corrected  the  photometry for  galactic
absorption.   From the  DIRBE/IRAS maps  of Schlegel  et  al.  (1998),
Kochanek   et   al.    (2002)   estimate  E(B-V)=0.464   towards   the
PKS~1830-211. We  correct our photometry  using this value and  a mean
galactic absorption  law with R$_V$=3.1 (Savage \&  Mathis, 1979). The
results are  indicated in Table 1.  Note that while  the correction is
reasonable for extragalactic objects,  it might be an over-estimate for
galactic objects such as object S1 and P.

%%%%%%%%%%%%%%%%%%%%%%%%%%%%%%%%%%%%%%%%%%%%%%%%%%%%%%%%%%%%%%%%%%%%%%

\section{Stars and galaxies towards PKS~1830--211 ?}

With  these  observations  of  improved  spatial  resolution,  greater
spectral  coverage  and  greater   depth,  it  is  easier  (but  still
challenging) to identify the different objects along the line of sight
to PKS~1830--211.

\subsection{Stars}

The  two   quasar  images  are  seen  simultaneously   with  a  decent
signal-to-noise only in  the $K$-band, i.e., on the  HST F205W images
and  on the  Gemini data.  The M-star,  observed  spectroscopically by
Djorgovski et~al.  (1992) and labeled S1 in Leh\'ar et~al.  (2000), is
seen  at all  wavelengths.   The total  field  contains also  numerous
galactic stars, given  the low galactic latitude of  the quasar, viz.,
$b = -5.7\deg$.

One point-like  object is seen between  the two quasar  images.  It is
labeled P  in Leh\'ar et~al.  (2000)  and in Figures  2-7, and remains
unresolved even in  the deep deconvolved F814W image,  in spite of the
high signal-to-noise.  The shallow depth of the Gemini images and the 
complicated PSF in the HST
images do not allow us to constrain the shape of object P.
We construct the color-magnitude diagram for 89 objects in the field
immediately surrounding PKS~1830--211 (Fig. \ref{fig7}), and find that
object~P falls in  the bulk of faint main-sequence  stars.  
This diagram has  to be taken with  caution since the
stars are  spread over a  broad range of distances,  metallicities and
reddening values.   Still, it is  indicative that object~P might  be a
galactic star seen on the line of sight to PKS~1830--211.

Star S1 was identified as an M dwarf by Djorgovski et al.  (1992) from
a Palomar 200-inch  spectrum.  From our photometry in  Table 1 (we use
here  the magnitude  not corrected  for galactic  reddening),  we have
m$_V$(S1) =  22.18 and m$_V$(P) =  26.20 with $(V-K)_{S1}$  = 5.46 and
$(V-K)_{P}$  = 5.13.  Such  colors  point towards  M  dwarf stars,  in
agreement with  Djorgovski et al.   (1992).  A value $V-K$  $\sim$ 5.3
corresponds  to an  M4 dwarf,  which  has a  absolute magnitude  $M_V$
$\sim$ 11.   If both S1  and P objects  were M4 dwarfs, they  would be
located,  ignoring any  reddening,  at a  distance  1.7 and  11.0~kpc,
respectively.  If we adopt 10 km\thinspace s$^{-1}$ as a typical
lower  limit for  the transverse  velocity, these  objects  would have
proper    motions    of    1~mas~yr$^{-1}$   and    0.2~mas~yr$^{-1}$,
respectively.  In WFPC2 frames, positions of stars have been measured
with  an  accuracy  of  about  0.02 pixel  (Anderson  \&  King  2000),
which corresponds to 1~mas on the PC of HST.
in the PC frames.  The HST observations in the F814W filter are split into
two epochs  separated by  529 days.  After carefully comparing these 
frames, neither S1 nor P show
significant proper motions. However, a time baseline of a few more years may
allow the  detection of some  proper motions, which would  confirm the
stellar character of these two objects.

\subsection{Galaxies}

Two more objects are visible between the two quasar images. One, 
labeled G in Leh\'ar et~al.  (2000), is seen without ambiguity
only in the F160W image. It is completely invisible in the F555W and
F814W images, and it is heavily contaminated by the PSF of nearby
point sources in the F205W image.  We are  able  to  measure  its F160W  
magnitude,  but
disagree with  the result of Leh\'ar  et~al.  (2000) by  more than one
magnitude.   Since there  is no  such  discrepancy for  all the  other
objects in the field, a  plausible explanation is contamination by PSF
residuals, acting in different ways  in our deconvolved images and the
PSF subtracted images  of Leh\'ar et~al.  (2000). Note  that there are
PSF residuals visible on the deconvolved F160W image. These are due to
differences between the PSF used  for the deconvolution and the actual
one of the  data. They are prominent near the  bright point sources S1
and QSO A.  For fainter sources such  as star P or QSO  B, they become
negligible, as they  are scaled down well below  the noise level. Note
that they are not visible close to  the faint stars to the West of the
field, hence even less near the fainter sources such as star P and QSO
B.  Object  G  can  therefore  be  considered  as real.  It  was
identified by  Leh\'ar et~al.  (2000)  as the one responsible  for the
molecular absorption lines observed at $z=0.89$ in the quasar spectra.

The F555W  and F814W  images allow  the detection of a second
new object, to the South of object P.  Only the brightest parts
of this object are visible in the shallow F814W data of Leh\'ar et~al.
(2000).  However, with the  new, much deeper exposures, two conspicuous
spiral arms appear on both sides of a brighter spot (Lens SP in Tables
1-2 and in Figures 2-6), which we identify as a possible position
for the bulge of an almost  face-on spiral.  The putative bulge of the
spiral is indicated by a circle  in the Figures 2-4 and labeled ``lens
SP''.  Note that given the lack  of resolution of the data, even after
deconvolution, the bulge  of galaxy SP is not  well defined. Therefore,
we consider  its center as the  barycenter of the area  defined by the
spiral  arms. It is  labeled Center  SA in  Table 2  and in  Figure 2.
Given  the uncertainties,  it is  one  of the  three possible  centers
considered in our  modeling. Using the F205W HST  data and the Gemini
data, we find that the extremely  red image B of the quasar is exactly
superposed with one  of the spiral arms, which  explains its faintness
in the optical.   This is clearly visible in  Figure \ref{fig2} but it
can  also  be seen  in  the  ``true-color''  images shown  in  Figures
\ref{fig5}  and  \ref{fig6},  which  help to  mentally  visualize  the
redshift information of the  data.  The first image (Fig.  \ref{fig5})
is  a composite  of  3 images,  through  the F814W,  F160W, and  F205W
filters.  It shows the two reddened quasar images and the central star
P,  which  has the  same  color as  the  field  stars.  Due  to
contamination by  the bright  quasar images, it  is difficult  in this
image to see clearly the two lensing galaxies.  Fig.  \ref{fig6} does not
use the F205W frame, where the  quasars are very bright. This image is
composed of the F814W image as  the blue channel, the F160W as the red
channel and the mean of the F814W and F160W as the green channel.  The
objects  between the  quasar  images  are now  easily  seen, with  the
face-on spiral  lens in blue  (note the blue  arm passing in  front of
quasar image B) and the much redder lens G, in green.

The spiral lens SP is very faint, but detected, in the F555W image. We
infer  a  $V-I$ color  of  0.8  $\pm$ 0.5  which  is  not well enough
constrained to set a first redshift estimate.  However, the very faint
$V$  and $I$-band  magnitudes  themselves indicate  a relatively  high
redshift,  given  the morphological  type  of  the  object.  This,  in
combination with the fact that the spiral arms are seen right in front
of  the heavily reddened  South-West quasar  component, makes  it very
likely that  we have discovered  the spiral  responsible for
the CO, HCO+, HCN and  HNC absorptions at $z=0.89$. We stress, however,
that we can  not completely exclude a much lower  redshift,
for example $z=0.19$, given the photometric error bars.

%%%%%%%%%%%%%%%%%%%%%%%%%%%%%%%%%%%%%%%%%%%%%%%%%%%%%%%%%%%%%%%%%%%%%%

\section{Modeling}

In spite of the complexity of the system, one can attempt to model the
different components  of the lens using the  few  available constraints
(position and flux ratio) and try to infer what might the correct lens
configuration       be.       We       have       considered      here 
Pseudo-Isothermal-Elliptical-Mass-Distributions  (PIEMD),  as used  for
example in Burud et al. (2002)  and Kneib et al. (1996).  Note however
that  the results quoted  here depend  only weakly  on the  exact mass
profile chosen. We  have considered the flux ratio  between the quasar
components to  be the one observed  in the radio,  unaffected by dust,
i.e.,   $F_A/F_B=1$.   We   chose  a   ``realistic''   cosmology  with
$H_0$=65~km\thinspace  s$^{-1}$\thinspace  Mpc$^{-1}$,  $\Lambda=0.7$,
and $\Omega_M=0.3$.

First, we  consider a single  mass distribution centered  at different
plausible locations:  galaxy G,  galaxy SP and  the barycenter  of the
spiral arms SA. Only for position G and SA, could we obtain a good fit
using an elliptical mass model with a predicted time-delay of about 40
days at position G and 31  days at position SA.  Changing the observed
flux ratio to  $F_A/F_B=1.5$ as measured by Lovell  et al. (1998) does
not significantly change the  predicted time-delay. Centering the mass
distribution  on star  P,  assuming it  is  the bulge  of the  lensing
galaxy,  predicts  time-delays  that  are  much higher  than  what  is
observed, for reasonable values of $H_0$.

Next, we consider two lens models with the lenses centered on G and  SP.
For simplicity, we assume G and SP to be at the same redshift, i.e.,
$z$=0.89, with  only the  velocity dispersion of  the two  galaxies as
a free parameter.  For circular mass distributions a good fit could 
not be found (reduced
$\chi^2=25$), and the mass of G was found to be 3 times larger than
SP.  We therefore added ellipticity.  We found that if $\epsilon \sim
0.2$ and PA$=-50$\degr\,  a good fit is achieved with  a mass ratio (G
over SP) of 3.5.  However, the  models never predict a time delay much
lower than 40 days, highly  incompatible with the observe one ($\Delta
t  = 26  \pm  5$  days). This  was  already pointed  out  in Lehar  et
al.  (2000),  who find  unrealistically  high  values  for the  Hubble
parameter if  the whole  lensing potential  were to be  due to  lens G
alone.  Changing  the  observed  flux  ratio  to
$F_A/F_B=1.5$ does not improve the situation.

Decreasing the redshifts  of the two lenses G and  SP to $z$=0.19 can also
be used to fit the data (with an elliptical  mass distribution for G),
but of course  predict very low time delays, of the  order of 10 days.
We have  therefore considered a model  composed of only  galaxy G (the
most massive  of the two)  and of the  large spiral galaxy seen  a few
arcsec to the  SW of PKS~1830-211, which will  introduce some external
shear.  As in  the model with two lenses (G+SP), no  decent fit can be
obtained  if G is  circular.  Good  fits are  found by  introducing an
ellipticity, with $\epsilon>0.3$, PA$\sim -90$\degr. 
In this case, the predicted time-delay ($\Delta t \sim 32 \pm 4$ days -
depending on the exact mass contribution of the lower redshift component)
matches the observations.  However,  we   reproduce  the  same  range  of
time-delay with a model that  consider only the lens G ($z$=0.89), but
located 0.08\arcsec\,  to the North  of its observed position.  Such a
shift is  large but  not out of  question given the  astrometric error
bars.

Constructing a model that  matches all the constraints (position, flux
ratio,  time delay)  is possible,  especially when  including  the low
redshift spiral galaxy to the  SW of the quasar.  
However,  in order to use  PKS~1830-211 to
constrain $H_0$, much better  observational information is required in
order to  characterise the  various mass component  along the  line of
sight to PKS~1830-211 and to ascertain the exact nature and reality of
all intervening objects.

%%%%%%%%%%%%%%%%%%%%%%%%%%%%%%%%%%%%%%%%%%%%%%%%%%%%%%%%%%%%%%%%%%%%%%

\section{Conclusion}

%The lensing potential in  PKS~1830-211 is composed of one face-on
%spiral galaxy with  a poorly defined center. If in  addition Lens G is
%real, also  with a poorly defined  center and possibly  at a different
%redshift, the  modeling of the  system becomes more  than challenging.
%Not  only we  might see  now two  lenses, but  very  few observational
%constraints   are  available.   For   example,  the   ellipticity  and
%orientation of  the lens  G are not  known. The  same is true  for the
%amplification  of  the quasar  images,  unknown  as  a consequence  of
%differential extinction by  the lenses. In case the  two lenses do not
%share the  same redshift one has  to deal with  the extra complication
%inherent to the  lensing effect of one lens on  the other. We conclude
%that there  is little  hope to successfully  model the system  with an
%accuracy comparable to that reached for other simpler systems.

The lensing potential in PKS~1830-211 is composed of one face-on
spiral galaxy with a poorly defined center and probably at z=0.89,
galaxy G also with a poorly defined center and unknown redshift and a
third galaxy, possibly at z=0.19.  With two and probably three galaxies
contributing to the lensing effect, we conclude that it will be difficult
to successfully model the system with an accuracy comparable to
that reached for other simpler systems.

Unless more observational  constraints become available, in particular
the precise redshifts and  positions of the lenses, PKS~1830--211 will
remain  of  little   interest  in  terms  of  modeling   and  for  the
determination of  $H_0$, even  though the time  delay is  now measured
with  good  accuracy. Never\-theless, PKS1830-211 is important for the  
study of absorption lines, microlensing
and absorption by dust.  

%Let us assume that the dust extinction is not
%at  $z=0.89$  but  at  $z=0.19$,  the estimates  of  the  differential
%extinction  done in Courbin  et~al.  (1998)  become much  higher, with
%$\Delta~E(B-V)=7.1$, assuming a  galactic extinction law applies, with
%$R_V=3.1$ (Savage  \& Mathis, 1979).   This is equivalent  to $A_V=17$
%magnitudes towards the faint quasar image.

PKS~1830-211  illustrates   very  well  the  need   for  high  spatial
resolution in  quasar lensing  studies.  Modeling of  this complicated
system will not  only require the redshift measurement  of the lenses,
but also the discovery of more lensed sources so that the lensing
potential can be probed in more than just the two points given by the
quasar images. The host galaxy of the  quasar can be used
to further  improve the  models, and arcs  or arclets might  appear at
very  high spatial  resolution.  Such data will only become available 
with the next generation of space
instruments or with AO systems mounted on large telescopes, possibly
coupled with integral field spectrographs.\\

{\sl Note: After the submission of this paper, Winn et al. (2002)
analysed some of the present data independently, and concluded that
star P is the bulge of one single lensing galaxy.  This hypothesis is
also plausible and modeling of the system is given by Winn et al.
(2002).}

%Photometric variations occur  in the SW image of  the quasar, and even
%changes  in   image  position  are  seen   at  millimeter  wavelengths
%(refs). Microlensing  is often invoked to explain  these changes. JP,
%Chris, do you  think we can say anything useful on  what can be learnt
%from microlensing in this object ?  Or anything else on the physics of
%the ISM from mm observations ?

%%%%%%%%%%%%%%%%%%%%%%%%%%%%%%%%%%%%%%%%%%%%%%%%%%%%%%%%%%%%%%%%%%%%%%


%% Included in this acknowledgments section are examples of the
%% AASTeX hypertext markup commands. Use \url without the optional [HREF]
%% argument when you want to print the url directly in the text. Otherwise,
%% use either \url or \anchor, with the HREF as the first argument and the
%% text to be printed in the second.

\acknowledgments

The multi-band HST archives data used in this paper were obtained
by the CfA-Arizona Space  Telescope LEns Survey (CASTLES).  We warmly
thank  Roser Pello  for  her  help with  the  redshift
estimates,  and  Tommy Wiklind  for  useful discussions.  Fr\'ed\'eric
Courbin  is  supported  by  Chilean  grant FONDECYT  3990024,  by  the
European   Southern   Observatory,    and   by   Marie   Curie   grant
MCFI-2001-0242.   Two collaborative grants  between Chile  and France
are   also   gratefully   acknowledged:   ECOS/CONICYT   CU00U05   and
CNRS/CONICYT 8730.   Jean-Paul Kneib acknowledges  support from French
CNRS.
 
%%%%%%%%%%%%%%%%%%%%%%%%%%%%%%%%%%%%%%%%%%%%%%%%%%%%%%%%%%%%%%%%%%%%%%

\begin{table*}[h]
\begin{center}
\caption{HST  photometry of  the objects  along the  line of  sight to
PKS~1830--211.}
\vspace*{2mm}
\begin{tabular}{l l l l l}
\hline 
\hline 
  Object            &  F555W            &  F814W            &  F160W            &  F205W              \\
                    &  WFPC2            &  WFPC2            &   NIC2            &   NIC2              \\
\hline
QSO A               &  25.62 $\pm$ 0.10 &  22.10 $\pm$ 0.03 &  17.04 $\pm$ 0.03 &  15.47 $\pm$ 0.01   \\ 
                    &  [24.19]          &  [21.42]          &  [16.82]          &  [15.34]            \\ 
QSO B               &  $>$26.50         &  $>$ 25.70        &  22.82 $\pm$ 0.20 &  19.53 $\pm$ 0.03   \\
                    &  [$>$25.07]       &  [$>$25.02]       &  [22.60]          &  [19.40]            \\
Star S1 (M-type)    &  22.18 $\pm$ 0.05 &  19.43 $\pm$ 0.02 &  17.20 $\pm$ 0.03 &  16.72 $\pm$ 0.03   \\
                    &  [20.75]          &  [18.75]          &  [16.98]          &  [16.59]            \\
Star P              &  26.20 $\pm$ 0.30 &  23.72 $\pm$ 0.10 &  21.72 $\pm$ 0.20 &  21.07 $\pm$ 0.20   \\
                    &  [24.77]          &  [23.04]          &  [21.50]          &  [20.94]            \\
Lens G              &   ...             &   ...             &  21.12 $\pm$ 0.50 &   ...               \\
                    &   ...             &   ...             &  [20.90]          &   ...               \\
Lens SP             &  26.64 $\pm$ 0.40 &  25.10 $\pm$ 0.20 &   ...             &   ...               \\
                    &  [25.21]          &  [24.42]          &   ...             &   ...               \\ 
Low redshift lens   &  23.71 $\pm$ 0.05 &  22.10 $\pm$ 0.02 &  20.25 $\pm$ 0.05 &   19.55 $\pm$ 0.05  \\
  (z=0.19 ?)        &  [22.28]          &  [21.42]          &  [20.03]          &   [19.42]           \\ 
\hline 
\end{tabular}
\tablecomments{The magnitudes given for Lens G, Lens SP and the
low redshift galaxy to the SW of PKS~1830-211 are measured on the
deconvolved images, through apertures of 0.5\arcsec\, in
diameter. Only the bulge of Lens SP is considered, the spiral arms
are invisible in the F555W image. Magnitudes between brackets are
corrected for galactic extinction.} 
\end{center}
\end{table*}

%%%%%%%%%%%%%%%%%%%%%%%%%%%%%%%%%%%%%%%%%%%%%%%%%%%%%%%%%%%%%%%%%%%%%%

%%%%%%%%%%%%%%%%%%%%%%%%%%%%%%%%%%%%%%%%%%%%%%%%%%%%%%%%%%%%%%%%%%%%%%

\begin{table*}[h]
\begin{center}
\caption{Astrometry  of  the  objects  along  the  line  of  sight  to
PKS~1830--211.}
\vspace*{2mm}
\begin{tabular}{l c c}
\hline 
\hline 
  Object                    &  X (\arcsec)          &  Y(\arcsec)       \\
\hline
  QSO A                     &    0.0000            &    0.0000             \\ 
  QSO B                     & $+$0.654 $\pm$ 0.002 & $-$0.725 $\pm$ 0.002  \\
  Star S1 (M-type)          & $-$0.091 $\pm$ 0.002 & $+$0.525 $\pm$ 0.002  \\
  Star P                    & $+$0.327 $\pm$ 0.004 & $-$0.491 $\pm$ 0.004  \\
Lens G                      & $+$0.519 $\pm$ 0.080 & $-$0.511 $\pm$ 0.080  \\
Lens SP                     & $+$0.285 $\pm$ 0.040 & $-$0.722 $\pm$ 0.040  \\ 
Center SA for spiral lens   & $+$0.300 $\pm$ 0.050 & $-$0.610 $\pm$ 0.050  \\ 
Low redshift lens (z=0.19 ?)  & $-$0.245 $\pm$ 0.050 & $-$2.490 $\pm$ 0.050  \\  
\hline
\end{tabular}
\tablecomments{As the different objects have very different colors,
the astrometry is obtained from the best available filter and then
matched to the coordinate system of the F814W image.  In particular,
QSO B is measured in the F205W image and in the Gemini $K$-band image.
Star~P, star S1 and the lens SP are measured in the F814W data. Lens
G is measured from the F160W data.}
\end{center}
\end{table*}

\clearpage

%%%%%%%%%%%%%%%%%%%%%%%%%%%%%%%%%%%%%%%%%%%%%%%%%%%%%%%%%%%%%%%%%%%%%%


\begin{figure}
%\plotone{fig01.ps}
\caption{Gemini North + Hokupa'a $K$-band image of the field surrounding
PKS~1830-211.  The guide star used for the wavefront measurement is
indicated.  The image quality on this 30-minute exposure is
0.14\arcsec.  This is twice the diffraction limit, but is still twice
as good as the HST in this band.  The two quasar images and the M-type
star S1 are already well separated on this un-deconvolved image.
\label{fig1}}
\end{figure}

\begin{figure}
%\plotone{fig02.ps}
\caption{{\bf Left:} Part of the HST/WFPC2 image of PKS~1830-211. This
image is a combination of all  available frames in the F814W filter (8
in  total), for  a  total exposure  time  of 6,400  sec. {\bf  Right:}
Simultaneous  deconvolution of  the 8  frames (see  text),  reaching a
resolution   of   0.046\arcsec,   and   sampled  with   a   pixel   of
0.023\arcsec. The various positions of  the two probable lenses and of
the very obscured (and hence  invisible here) QSO~B are indicated with
circles. Note the obvious spiral shape between the quasar images, with
one  spiral arm  passing right  onto the  line of  sight to  QSO~B. We
identify  this  galaxy  as   the  probable  source  of  absorption  at
$z=0.89$. The center  of the ellipse shown in  this figure defines the
barycenter  SA of  the spiral  (see  main text).  It is  one of  three
centers, together with SP and G, considered in the modelling. There is
no significant trace  in this image of the much redder  lens G, but we
plot its position as measured from the near-IR F160W image, in overlay
on the F814W image.
\label{fig2}}
\end{figure}

\begin{figure}
%\plotone{fig03.ps}
\caption{{\bf  Left:} Part  of the  HST/NICMOS2 image  of PKS~1830-211
obtained  in the  F160W  filter (4  individual  frames). {\bf  Right:}
Simultaneous  deconvolution of  the four  F160W images,  with  a final
resolution of 0.075\arcsec.  Note the object right to the West of star
P.  Lens G is visible in this frame only. There is no trace of the
spiral galaxy SP, which is marked with a circle.
\label{fig3}}
\end{figure}

\begin{figure}
%\plotone{fig04.ps}
\caption{{\bf Left:} Gemini + Hokupa'a $K$-band image of PKS~1830-211.
This image shows a zoom in  the same data as in Fig. \ref{fig1}.  {\bf
Right:} The simultaneous deconvolution of  8 stacks of images allows 
a resolution of  0.020\arcsec\, to be reached, 
sampled  on a  grid of  pixels of
0.010\arcsec.  The  data, too shallow to reveal  the lensing galaxies,
is useful  to measure the positions of the  quasar images, with a
very high precision. The point-like object~P (not modelled as a 
point source in the Gemini data) is clearly visible.
\label{fig4}}
\end{figure}

\begin{figure}
%\plotone{fig05.ps}
\caption{Composite image  obtained from the central 
parts of the  HST $I$ (F814W),
$H$ (F160W), and $K$  (F205W) band data.  The different objects identified
in the single  band images are identified. Note  the strong similarity
between the colors of star~P and the other stars in the field. The two
quasar images, extremely reddened, are prominent in the red channel of
the image  (i.e., the $K$-band image).   Lens SP is visible  as a blue
fuzz between the  quasar images, while lens G, best  seen in F160W, is
visible as  a greenish  object (F160W being  the green channel  of the
composite image).   Note the low-redshift  (?) galaxy to the  south of
the lensed quasar.
\label{fig5}}
\end{figure}

\begin{figure}
%\plotone{fig06.ps}
\caption{Same field  as in Figure \ref{fig5} but  constructed with the
$I$-band (F814W) data as the blue channel, the $H$-band (F160W) as the
red channel, and the mean of  the $I$ and $H$-band images as the green
channel.  Light  contamination by the quasar images  affects much less
the lens galaxies, which are clearly visible, either in blue (spiral lens
SP)  or green  (lens  G). Note  the  blue spiral  arm  passing at  the
position of  quasar image  B. The objects  marked ``??''  are probably
structures in the PSF of the $H$-band image.  They are cleanly removed in
the deconvolution in Fig. \ref{fig3}
\label{fig6}}
\end{figure}

\begin{figure}
\plotone{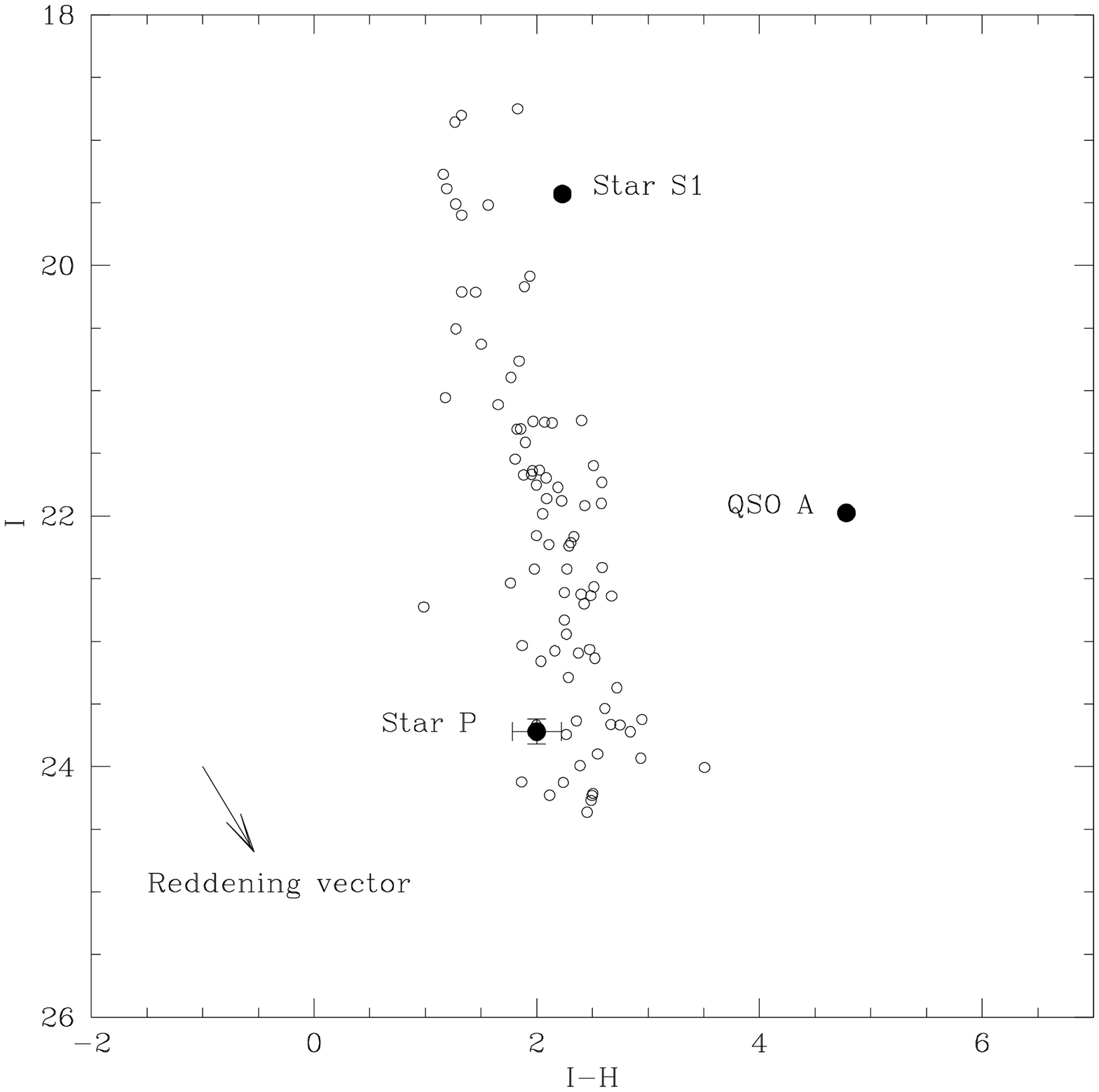}
\caption{   Color   magnitude    diagram   for   89   objects   around
PKS~1830--211. The stellar main sequence is seen. When looking at this
diagram, one should note that  stars towards the Galactic bulge are
spread over a broad  range of distances, metallicities and reddenings.
The open  dots show the stars.  The different objects  of interest are
ploted as larger black dots, such as QSO A, Star P and Star S1.  Star P
falls on the  stellar main sequence, within the  error bars.  The data
points are not de-reddened. The reddening vector is indicated.
\label{fig7}}
\end{figure}

\end{document}